\documentclass[12pt,preprint,showpacs,amsmath,amssymb]{revtex4-1}
\usepackage{mathrsfs}
\usepackage{slashed}
\usepackage{epsfig}
\usepackage{graphicx,color,multirow}
\usepackage{dcolumn}
\usepackage{bm}

\def\beal#1\eeal{\begin{align}#1\end{align}}

\newcommand{\bef}{\begin{figure}[htb]\centering}
\newcommand{\eef}{\end{figure}}
\newcommand{\beq}{\begin{equation}}
\newcommand{\eeq}{\end{equation}}
\newcommand{\bea}{\begin{eqnarray}}
\newcommand{\eea}{\end{eqnarray}}

\newcommand{\nm}{\nonumber}

\def\sk{\not{\hbox{\kern-2.1pt $k$}}}
\def\spg{\not{\hbox{\kern-2.1pt $p_g$}}}
\def\spt{\not{\hbox{\kern-1.7pt $p_t$}}}
\def\bmaT{\left(\begin{array}{ccc}}
\def\emaT{\end{array}\right)}
\let\jnfont=\rm
\def\NPB{\jnfont Nucl.\ Phys.\ B}
\def\PLB{\jnfont Phys.\ Lett.\ B}
\def\EPJC{\jnfont Euro.\ Phys.\ J.\ C}
\def\EPJP{\jnfont Euro.\ Phys.\ J.\ Plus}
\def\PRD{\jnfont Phys.\ Rev.\ D}
\def\PRL {\jnfont Phys.\ Rev.\ Lett.}

\def\CTP{\jnfont Commun.\ Theor.\ Phys.}
\def\JHEP{\jnfont J. High \ Ener.\  Phys.}

\def\PLB{\jnfont Phys. Lett. B}
\begin{document}

\title{  Top Rare Decays $t\to cV$ 
in the Mirror Twin Higgs Models }
%

\author{Su-Ya Bie$^{1}$ }
\author{Guo-Li Liu$^{1}$} \email{ guoliliu@zzu.edu.cn, corresponding author}
\author{Wenyu Wang$^{2}$}\email{wywang@bjut.edu.cn, corresponding author}
	\affiliation{$^{1}$School of Physics and Microelectronics, Zhengzhou University, Zhengzhou 450001, P. R. China,}
    \affiliation{$^{2}$School of Physics, Beijing University of Technology, Beijing, P.R.China}
\begin{abstract}
The decay $t \to c V $ ($V=\gamma,~Z,~g$) process in the mirror twin Higgs models with the colorless top partners are studied in this paper.
We found that the branching ratios of these decays can in some parameter spaces alter the standard model expectations greatly
and may be detectable according to the currently precision electroweak measurements.
Thus, the constraints on the model parameters may be obtained from the branching fraction of the decay processes,
which may serve as a robust detection to this new physics model.
\end{abstract}

\maketitle

\newpage
\section{Introduction}
In high energy physics today, one of the most important questions is to determine whether the Higgs mass tuning indeed
exists in nature or whether the electroweak scale is set by a mechanism that does not need
a large cancellation. This is the issue of the Higgs naturalness, or the hierarchy problem.

An attractive dynamical kind model to solve the hierarchy problem is to introduce a new symmetry
which protects the Higgs against large radiative corrections. That is, these models invoke such a symmetry
that implies the particles existing beyond the standard model (SM) which consists of the "symmetry partners"
of known SM fields.

We know that the hierarchy problem depends on the top quark one loop diagram,
therefore ANY model that resolves the hierarchy problem must introduce the top quark
symmetry partners, the "top partners".
To avoid significant residual tuning, on the other hand, the top partners are expected to
have masses at or below the TeV scale.
For examples, in supersymmetric models (for a review see  \cite{9709356}), in little
Higgs models \cite{0105239,0206020,0206021,0407143} (for a review see \cite{0502182}), there exist the scalar stops
and vector-like fermionic top-primes as top partners, respectively.
In these models the new symmetry protects the Higgs commuting with the SM gauge symmetries,
so the quantum numbers of the top partners are identical as those of the top quark.

The searches for these colored top partners, both scalar and fermionic, however, have so far suffered
stringent limits from the large hadron collider (LHC) searches(see e.g. \cite{1308.1586,CMS_Col,1407.0583,1406.1122}),
 so theories which include colorless top partners,
i.e, not charged under the strong interactions, appear ever more compelling.
Since the production cross sections of uncolored top partners are many orders of magnitude smaller
than in the colored case, this allows a simple understanding of why these particles have so far
escaped discovery.

Colorless top partners occur in scenarios where the symmetry is district, but not the global symmetry
as in little Higgs theories \cite{0506256,0609152,0812.0843}.
 By far, the most striking possibility of the uncolored top partners
is the mirror twin Higgs (MTH) model, where the Higgs is protected by the discrete $Z_2$ subgroup\cite{0506256}
(see also \cite{0812.0843,0510273,0604076,1312.1341,1410.6808}).


On the other side, the huge mass of the top quark makes its lifetime very short,
and it decays without non-perturbative hadronization effects.
So there is still room for non-standard top quark interactions, such as the productions and the decays.
Moreover, the top quark strongly interacts with the yet-mysterious Higgs boson.
So the detailed studies of top-quark interactions would be useful to explore the mechanism of electroweak symmetry breaking
and some properties of the Higgs boson.


In the SM, the flavor changing neutral currents (FCNCs) are absent at the tree-level, and in the loop-level, they are
strongly suppressed by the Glashow-Iliopoulos-Maiani (GIM) mechanism.
Within the SM, the Decays of the top quark induced by FCNC interactions are known to be extremely rare.
Thus, the FCNC interactions are of utmost importance in constraining the beyond SM (BSM) physics.
However, this kind of loop-driven processes can get contributions from new physics particles and new couplings
and alter greatly the SM predictions for these processes.
In the present paper, we will consider the rare decays of $t\to cV$ ($V=\gamma,~Z,~g$)
in the context of mirror twin Higgs(MTH) models.

The paper is organized as follows. In Section 2, we briefly describe the realization of the MTH models
with colorless top partners and introduce the related couplings with the top rare decay.
Section 3 is dedicated to discussions about the calculation of the the three decay processes in these
models. The results are  elaborated in Section 4. Section 5 will list the compatibility of parameter space
with phenomenological constraints coming from electroweak precision
data, LHC observations. Finally, the summarization and conclusion are given in Section 6.

\section{the realization of the mirror twin Higgs models with colorless top partners and the correlative couplings\cite{1411.3310}}

\subsection{The Model and Cancellation Mechanism\label{ssec:twinmodel}}

In the MTH models, there assumes to exist a $Z_2$ distinct symmetry
which exchanges the complete SM with a mirror copy of the SM, and this copy is called
the twin sector. 
Besides this, the global symmetry of the Higgs sector of the theory is approximate,
which may be taken as either SU(4)$\times$U(1) or O(8), and the SM
Yukawa couplings, and the SM electroweak gauge interactions will explicitly violate the
global symmetry.
The gauge subgroup contains the SU(2)$\times$U(1) electroweak interactions
of the SM and of the twin sector.
After the global symmetry is spontaneously broken, the SM Higgs doublet emerges as light
pseudo-Nambu-Goldstone boson.
In spite that the global symmetry will be violated, the discrete $Z_2$
symmetry, however, will be exact to ensure the absence of quadratically divergent contributions to
the Higgs mass at the one-loop level.

In the following SU(4)$\times$U(1) group is taken as an example of the global symmetry to describe
the cancellation of the quadratic divergences in this model, and
the gauge subgroup of the SM and twin sectors can be taken as SU(2)$\times$SU(2)$\times$U(1) and $U(1)$, respectively.
Labels $A$ and $B$ will be used to denote the SM and twin sectors, and under the action of the discrete $Z_2$
symmetry, the labels $A$ and $B$ will exchange with each other, $A \leftrightarrow B$.
 Then, The field $H$, which transforms as the fundamental representation under the global SU(4) symmetry, can be written as
 \begin{equation}
H=\left(\begin{array}{c}
H_A\\
H_B
\end{array} \right) \; ,
 \end{equation}
where $H_A$ and $H_B $ represent the SM Higgs doublet and the twin doublet, respectively.

The SU(4) potential for $H$ is
 \begin{equation}
m^2 H^{\dagger}H + \lambda (H^{\dagger} H)^2
 \end{equation}
When the parameter $m^2$ is negative, the global symmetry is
spontaneously broken, SU(4)$\times$U(1)$\to$ SU(3)$\times$U(1),
and thus the gauge and Yukawa interactions engender radiative corrections
which violate the global symmetry and generate a mass for $H_A$.

To cancel the quadratically divergent corrections,
the top Yukawa coupling can be taken as,
  \begin{equation}
  \lambda_{Ai} H_A q_{Ai} t_A + \lambda_{Bi} H_B q_{Bi} t_B \,.
\label{eq:twntopsect}
 \end{equation}
Due to the $Z_2$ symmetry, $\lambda_{At}=\lambda_{Bt}=\lambda$ ,
so that, at one loop order, the quadratically divergent corrections to the
Higgs potential can be generated and cancelled out by these interactions.
The corrections are ($\Lambda$ is the ultraviolet (UV) cutoff.)
 \begin{equation}
\Delta V = \frac{3}{8 \pi^2} \Lambda^2 \left(
\lambda_{At}^2 H_A^{\dag} H_A + \lambda_{Bt}^2 H_B^{\dag} H_B \right) = \frac{3 \lambda^2}{8 \pi^2} \Lambda^2 H^{\dagger} H \;\; ,
 \end{equation}

Thus, with the $Z_2$ symmetry, the contribution above respects the global symmetry,
so it cannot contribute to the mass of the Nambu-Goldstone bosons.

More generally, the cancellation mechanism of the Higgs mass can also be understood in the low
effective theory. $H$ can then be written as,
\begin{equation}
H=\left(\begin{array}{c}
H_A\\
H_B
\end{array} \right) =
\exp\left(\frac{i}{f}\Pi \right)\left(\begin{array}{c}
0\\
0\\
0\\
f
\end{array} \right) \; .
 \end{equation}
Where $f$ is the symmetry breaking vacuum expectation value(VEV),
and $\Pi$ is
\begin{equation}
\Pi=\left(\begin{array}{ccc|c}
0&0&0&h_1\\
0&0&0&h_2\\
0&0&0&0\\ \hline
h_1^{\ast}&h_2^{\ast}&0&0
\end{array} \right).
 \end{equation}
Expanding out the exponential, we have
 \begin{equation}
H=\left(\begin{array}{c}
\displaystyle \bm{h}\frac{if}{\sqrt{\bm{h}^{\dag}\bm{h}}}\sin\left( \frac{\sqrt{\bm{h}^{\dag}\bm{h}}}{f} \right)\\
0\\
\displaystyle f\cos\left( \frac{\sqrt{\bm{h}^{\dag}\bm{h}}}{f} \right)
\end{array} \right)
\end{equation}
 where $\bm{h} = (h_1, h_2)^{\text{T}}$ is the Higgs doublet of the SM
 \begin{align}
\displaystyle H_A&= \bm{h}\frac{if}{\sqrt{\bm{h}^{\dag}\bm{h}}}\sin\left( \frac{\sqrt{\bm{h}^{\dag}\bm{h}}}{f} \right)=i \bm{h}+\ldots\,, \\
 H_B&= \left(\begin{array}{c}
0\\
f\cos\left( \frac{\sqrt{\bm{h}^{\dag}\bm{h}}}{f} \right)
\end{array} \right)=\left(\begin{array}{c}
0\\
\displaystyle f-\frac{1}{2f}\bm{h}^{\dag}\bm{h}+\ldots
\end{array} \right).
 \end{align}
 Now considering Eq.~\ref{eq:twntopsect} in quadratic order of $\bm{h}$,
 \begin{equation}
i\lambda_i\bm{h} q_{Ai}t_A+\lambda_i\left(f-\frac{1}{2f}\bm{h}^{\dag}\bm{h} \right) q_{Bi}t_B \; . \label{eq:twntoplag}
 \end{equation}
 Thus the quadratic divergence arising from the first diagram is exactly canceled by that of the second
 via evaluating these contributions.


\subsection{The quark flavor changing couplings\label{ssec:fcc}}
Now we focus on the flavor changing of the top quark. Firstly,
we determine the low energy couplings of the Higgs.
Choosing unitary gauge in the visible sector with $h_1=0$ and $h_2=(v+\rho)/\sqrt{2}$, we obtain
 \begin{equation}\begin{array}{cc}
H_A=\left(\begin{array}{c}
0\\
\displaystyle if\sin\left(\frac{v+\rho}{\sqrt{2}f} \right)
\end{array}\right), &H_B=\left(\begin{array}{c}
0\\
\displaystyle f\cos\left(\frac{v+\rho}{\sqrt{2}f} \right)
\end{array}\right).
\end{array}
 \end{equation}
 The kinetic terms are
 \begin{equation}
\left|D_{\mu}^AH_A \right|^2+\left|D_{\mu}^BH_B \right|^2
 \end{equation}
 where the $D^{A,B}$ denote the covariant derivative of the $A,B$ gauge bosons.
 From the above equation one can obtain the masses of the $W^\pm$ and $Z$ gauge bosons in
the visible $A$ and twin sectors $B$ and their couplings to the Higgs, $\rho$, which
determine the relation of the Higgs SM VEV $v_{\text{EW}}=$246 GeV and the MTH parameters $v$ and $f$,
 \begin{equation}
v_{\text{EW}}=\sqrt{2}f\sin\left(\frac{v}{\sqrt{2}f} \right)\equiv\sqrt{2}f\sin\vartheta  \; .
\label{vew_v}
\end{equation}
where the angle $\vartheta=\frac{v}{\sqrt{2}f}$, and when $v=v_{\text{EW}}$, $v \ll f$, or equivalently
$\vartheta\ll 1$.

Expanding  the top quark sector \eqref{eq:twntopsect} in the unitary gauge,
\begin{align} \nm
&\lambda_i\left[ifq_{Ai}t_A\sin\left(\frac{v+\rho}{\sqrt{2}f} \right) +fq_{Bi}t_B\cos\left( \frac{v+\rho}{\sqrt{2}f}\right) \right]\\
=&i\frac{\lambda_i v_{\text{EW}}} {\sqrt{2}}q_{Ai}t_A\left[1+\frac{\rho}{v_{\text{EW}}}\cos\vartheta \right]+ \lambda_i fq_{Bi}t_B\cos\vartheta\left[ 1-\frac{\rho}{v_{\text{EW}}} \tan\vartheta\sin\vartheta \right]
\label{tc_coup}
\end{align}
Here $q_i$ can be quarks $u,~c$ or $t$.

 From this, we can also see that the mass of the top quark's mirror twin partner is
\begin{equation}
m_T=\lambda_t f\cos\vartheta=m_t\cot\vartheta\, .
\end{equation}


From Eq.(\ref{tc_coup}), we can also see clearly that the scalar $\rho$ acts
as the SM-like Higgs, and it consists of both visible and invisible
part in some ratio according to a certain parameters.

\section{Calculation of the top rare decays $t\to cV$}
\begin{figure}[!htbp]
  \centering
   \includegraphics[scale=0.95]{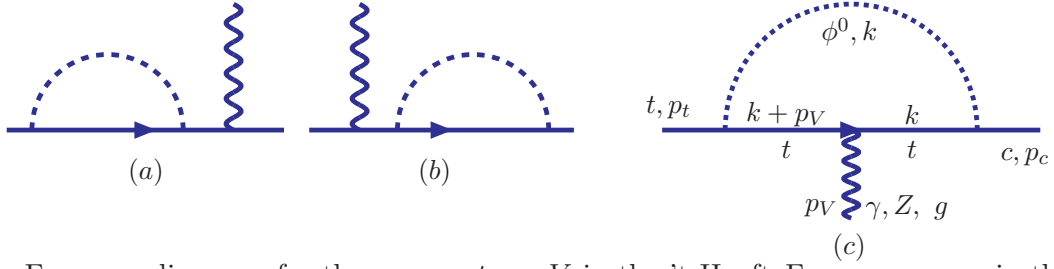}
  \vspace{-0.9cm}\caption[]{Feynman diagrams for the process $t\to cV$ in the 't Hooft--Feynman gauge in the MTH model. }
\label{fig:tcv}
\end{figure}

From Eq.(\ref{tc_coup}), we can see the flavor changing couplings mediated by the neutral scalar,
so the FCNC decays $t\to c V$ can be realized by it, and the Feynman diagrams are listed in Fig.\ref{fig:tcv}.
We can see from Eq.(\ref{tc_coup}) that the structure the fermions couplings to the scalar $\rho$ are very simple.
To be more general, we write the couplings of the scalar to the fermions as
\beq
\rho q_{Ai}\bar t_A: i\frac{\lambda_i} {\sqrt{2}} (c+d\gamma^5)
\label{tc_coup_gene}
\eeq
where $q_{Ai}=t,~c$(we only consider the visible section A ), $\lambda_c=V_{tc}\lambda_t$ and $V_{tc}$ is the
ratio of the two couplings, a bit like the CKM matrix element $U_{tc}$.
The parameters $c$ and $d$ are just to expand the lorentz structure,
and from Eq. (\ref{tc_coup}), we can see that $c=\frac{1}{2},~ d=0$.
In the following discussion, however, we can also release the constricts and to check
the influence on the branching ratios in Fig. \ref{fig3}.

\subsection{The amplitude and the width of $t\to cg$, $t\to c \gamma$ and $t\to cZ$}

With the general coupling of the Yukawa form, we can write out the amplitude of the decay $t\to cg$, (taking
Fig.\ref{fig:tcv} (c) as an example)
\bea \nm
{\cal M}_c&=& \bar u_c (i\frac{\lambda_c } {\sqrt{2}} )(c+d\gamma^5)\frac{i}{\sk-m_t}(-ig_s\gamma^\mu T^a)
\frac{i}{\sk+\spg-m_t}(-i\frac{\lambda_t } {\sqrt{2}} )    \\ \nm &&
(c-d\gamma^5) \frac{i}{(k+p_g-p_t)^2-m^2_\rho} u_t \epsilon_\mu \\ \nm
&=&-\frac{\lambda_c\lambda_t }{2}g_sT^a\frac{1}{k^2-m_t^2}\frac{1}{(k+p_g)^2-m_t^2}\frac{1}{(k+p_g-p_t)^2-m^2_\rho} \\ \nm &&
\bar u_c  (c+d\gamma^5)(\sk+m_t)\gamma^\mu(\sk+\spg+m_t)(c-d\gamma^5)u_t \epsilon_\mu
\eea
Note that in the above formula, we have omitted the common integrated factor $\tau^{2\epsilon }\int^{+\infty}_{-\infty} \frac{d^nk}{(2\pi)^n}$,
$\epsilon=1-\frac{n}{2}$, $\tau$ is the scale factor introduced to keep the dimension of the coupling constants unchanged.
 The decay width in general is given by
\beq
\Gamma_{t\to cg} = C_F \frac{1}{32\pi}|M|^2
\eeq  
where $C_{F} = 4/3$ is a color factor.

As for the width of $t\to c\gamma$, we can simply obtain it by replacing the coupling $g_s$ with $e$ in the amplitude, and $C_{F}$ with $1$.
Since the coupling of $Z\bar t t$ is a little complicated than that of $\gamma \bar t t $, so in the calculation of the
$t\to c Z$, we need to replace the $t\to c\gamma$ parameters $e$ with $e/(2s_Wc_W)$,
and the structure $\gamma^\mu$ with $\gamma^\mu(P_L-2s_W^2)$.

In this scalar-mediated decay process, the sum of the one-loop divergent terms is zero.
That's to say, the one-loop divergences cancel out with each other in the Feynman gauge, 
so we can safely use the calculating tool of LoopTools\cite{looptools}.

Of course, the effective vertex $\bar{t}cV$ is a 4-component Lorentz vector and also
a $4 \times 4$ matrix in Dirac space and need to be managed.
One trick is that the tensor loop functions are retained rather than expanding them
in term of scalar loop functions as usual \cite{Hooft}.

In its realization of Fortran, a dimension-three array $V(i,j,k)$
with $i$ (=1,2,3,4) labelling the Lorentz index and  $j,k$ (=1,2,3,4)
labelling the spinor indices are used. See the details in Ref.\cite{0702264}.

\subsection{The top total width and the upper bounds to the rare decays }

 The decay width of the dominant decay mode of the top
quark  $t\to bW$ is given by ~\cite{1807.01481,Dey:2016cve}
\begin{align}
\Gamma_{t\to bW} = \frac{G_{F}}{8\sqrt{2}\pi}|V_{tb}|^{2}
                    m_{t}^{3}
                    \left[
                    1 - 3\left(\frac{m_{W}}{m_{t}}\right)^{4}
                      + 2\left(\frac{m_{W}}{m_{t}}\right)^{6}
                    \right].
\end{align}
The above equation gives $\Gamma_{t\to bW} \sim 1.5$~GeV.
So the branching ratio of any other mode $t\to X$ is
\begin{align}
\text{BR}(t\to X) = \frac{\Gamma_{t\to X}}{\Gamma_{t\to bW}}.
\end{align}
The SM predictions for the $t\to cg$, $t \to c\gamma$ and $t\to cZ$ branching ratios are \cite{1807.01481,Dey:2016cve,Abbas:2015cua}
\beq
\text{BR}(t\to cg) = \left(4.6^{+1.1}_{-0.9}     \pm 0.4^{+2.1}_{-0.7}   \right)\times 10^{-12},
\label{tcg_sm}
\eeq
\beq
BR(t \to c\gamma) = \left(4.6^{+1.2}_{-1.0}\pm 0.4^{+1.6}_{-0.5} \right)\times 10^{-14},
\label{tcr_sm}
\eeq
\beq
\text{BR}(t\to cZ) =(1.03\pm 0.06)\times 10^{-14},
\label{tcz_sm}
\eeq
LHC has searched for these rare decays and give their upper bounds~\cite{lhcTopWGNov:2017,Aad:2015gea, Khachatryan:2016sib, ATLAS-CONF-2017-070,Sirunyan:2017kkr, t_cr_bounds}:
\beq
 BR(t\to cg) < 2\times 10^{-4}.
 \label{tcg_lhc}
 \eeq
\beq
 BR(t\to cZ) < 2\times 10^{-4}.
 \label{tcr_lhc}
 \eeq
\beq
 Br(t \to c\gamma)< 1.82\times 10^{-3}.
\label{tcz_lhc}
\eeq
From Eq.(\ref{tcg_lhc}) to Eq.(\ref{tcz_lhc}), we can see that the predictions of the
 SM Eq.(\ref{tcg_sm}) to Eq.(\ref{tcz_sm}), are impossible to be probed
at the LHC, so any signal of the flavor decays of this kind is inevertably the signature
of the new physics.

Some BSM scenarios may predict an enhanced branching ratio of these rare modes up to the level that can be detected
in the future colliders, such as 2HDM~\cite{Eilam:1990zc, Abbas:2015cua, Gaitan:2017tka},
     left-right symmetric model~\cite{Gaitan:2004by}, MSSM~\cite{0702264},
     $R$-parity violating SUSY~\cite{Bardhan:2016txk}, warped extra dimensional
     models~\cite{Agashe:2006wa, Gao:2013fxa},  UED models~\cite{GonzalezSprinberg:2007zz}, mUED and nmUED models \cite{Dey:2016cve,1807.01481}
     and composite Higgs model~\cite{Agashe:2009di,tc2-comp-fcnc}, etc.
In Ref.~\cite{CorderoCid:2004vi, Datta:2009zb} one can find that the effective Lagrangian approach is used to study of rare top decays.
Other collider stuides to the search of these rare decays can be found in Ref. \cite{DiazCruz:1989ub, Mele:1998ag, AguilarSaavedra:2002ns, AguilarSaavedra:2004wm, Chen:2013qta, Khanpour:2014xla, Hesari:2014eua, Kim:2015oua, Hesari:2015oya, Khatibi:2015aal, Malekhosseini:2018fgp, Banerjee:2018fsx}, etc.
In the following, we will check whether the prediction of the MTH models may arrive at the detectable level
and provide constraints to the model parameters.

\subsection{The results for $t\to cV$ }
\def\figsubcap#1{\par\noindent\centering\footnotesize(#1)}
 \begin{figure}[htb]%
 \begin{center}       \hspace{-1.5cm}
  \parbox{5cm}{\epsfig{figure=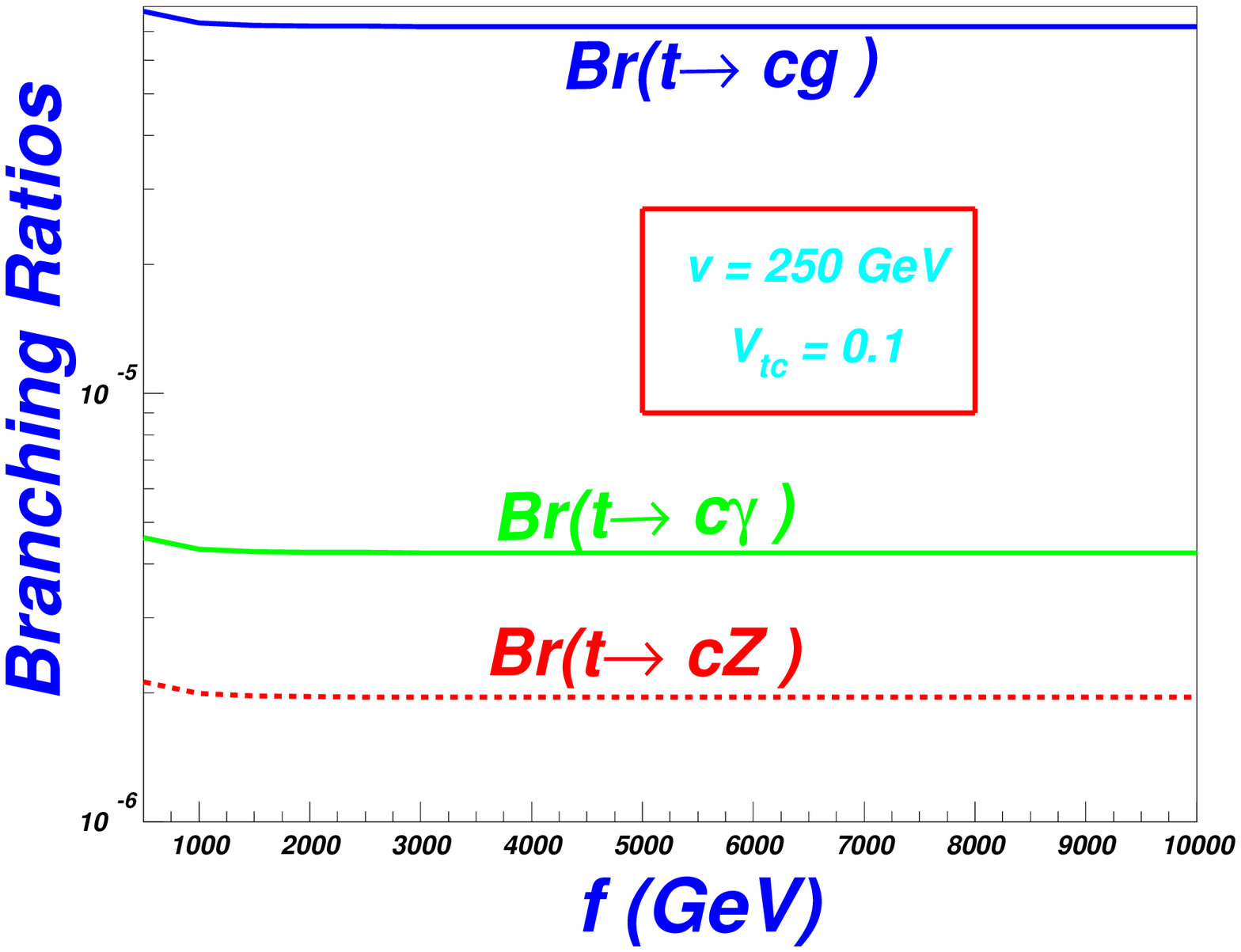,width=5cm}    \figsubcap{a}}
  \parbox{5cm}{\epsfig{figure=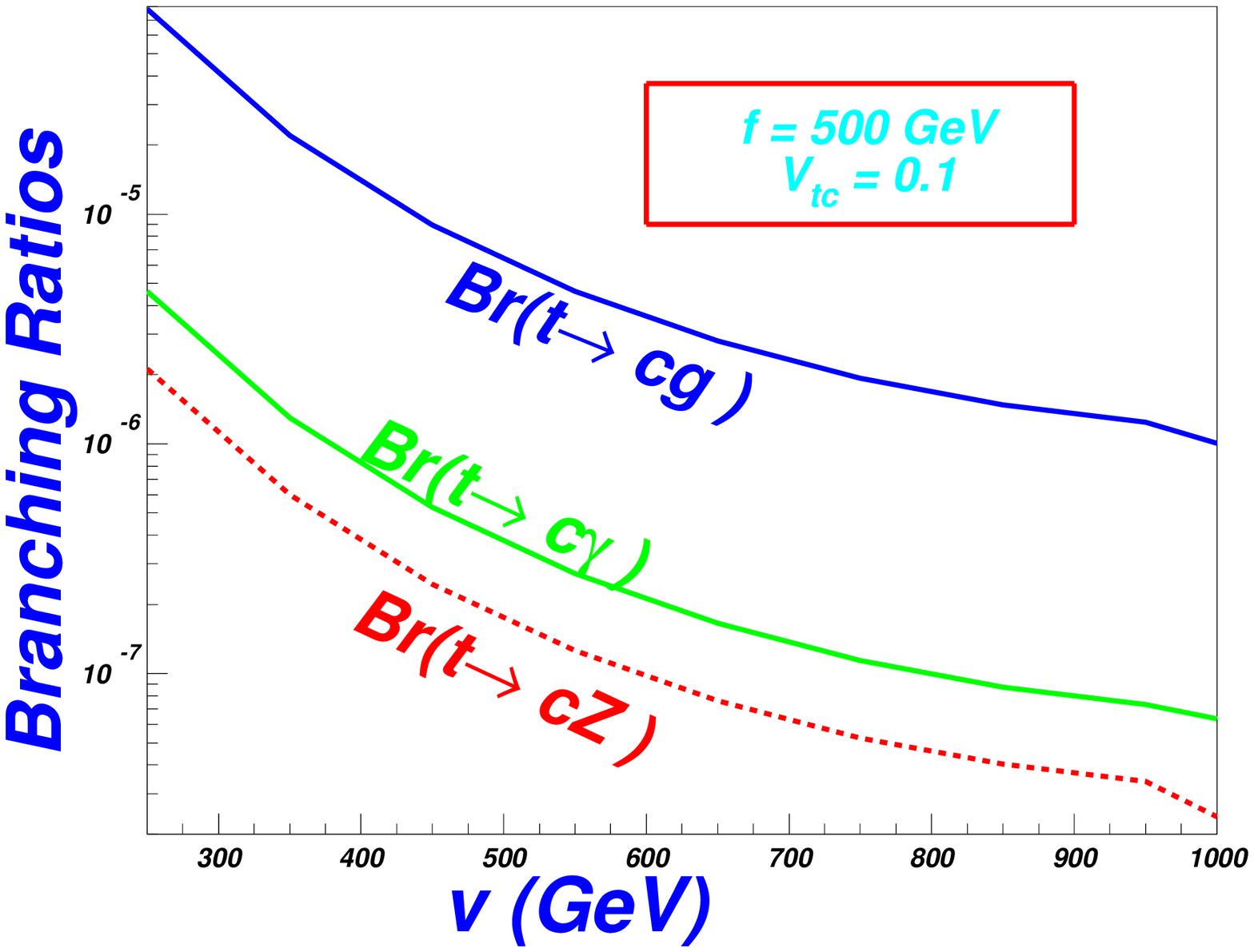,width=5cm}    \figsubcap{b}}
  \parbox{5cm}{\epsfig{figure=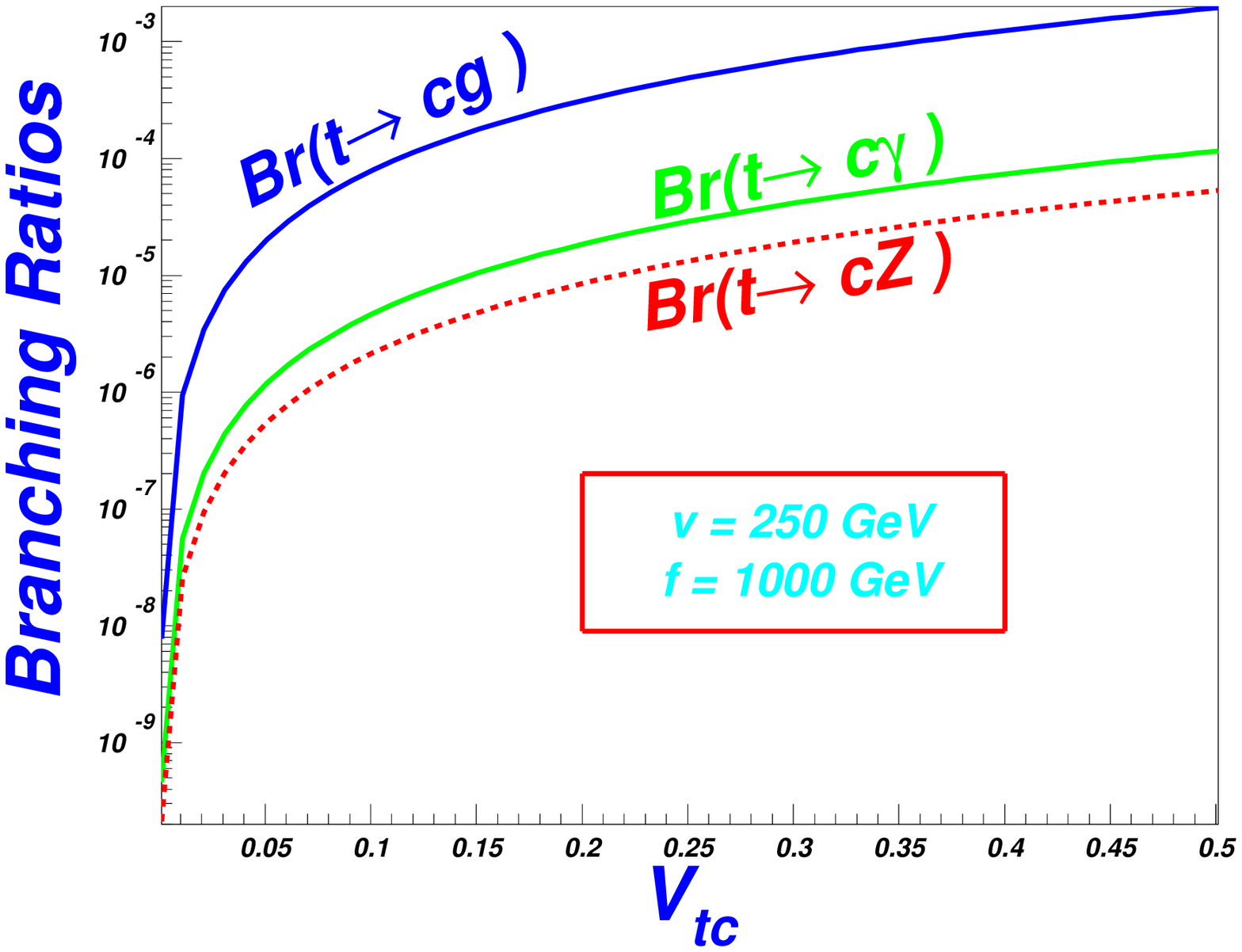,width=5cm}   \figsubcap{c}}
 \caption{ The one-loop level branching ratios of the three processes in the MTH model.
  \label{fig2} }
   \end{center}
 \end{figure}

In order to find constraints on the parameters $f$, the VEV $v$ (To find the relation between it and the electroweak
VEV $v_{EW}$, one can refer to Eq.(\ref{vew_v})), and the flavor changing coupling $V_{tc}$,
the one-loop level branching ratios of the three processes
are given in Fig.\ref{fig2}.
We take $m_t = 172.5$ GeV and $M_Z = 91.2$ GeV,
 and other physical constants are obtained from the Review of Particle Physics \cite{pdg-2018}.

Figures in Fig.\ref{fig2} show that the branching ratios are respectively in the range: $Br(t\to cg) \sim  10^{-3}-10^{-6}$,
$Br(t\to c\gamma) \sim 10^{-4}-10^{-8}$ and $Br(t\to cZ) \sim 10^{-5}-10^{-8}$.
The reason that the $t\to cg$ is larger than those of the $t\to c\gamma~/Z$
is that the $g\bar q q$ coupling in this process with QCD coupling $\alpha_S$ involved,
 is one-order larger than the electroweak coupling $\gamma \bar q q \sim \alpha_e$.

We can see from Fig.\ref{fig2} that the influence of the parameters on the branching ratios is not the same:
Varying $f$ or $v$ doesn't make much difference when other parameters are fixed, as shown in Fig.\ref{fig2}(a)(b),
while varying $V_{tc}$ will change the branching ratios largely. However, the alteration is an associative effect of
the parameters $f$, $v$ and $V_{tc}$, since in the couplings Eq.(\ref{tc_coup_gene}), $\lambda_c$ is connected with
the threes and so not quadratically increasing with the increasing $V_{tc}$.

From Fig.\ref{fig2}, we can also see that the process $t\to cg$ may be hopeful to arrive at the detectable level
 according to the experimental bounds in Eqs. (\ref{tcg_lhc}), (\ref{tcr_lhc}), (\ref{tcz_lhc}),
 and may provide constraints on the parameters. So in the following, we will only consider the process $t\to cg$.

\begin{figure}[!htbp]
  \centering
    \includegraphics[scale=0.5]{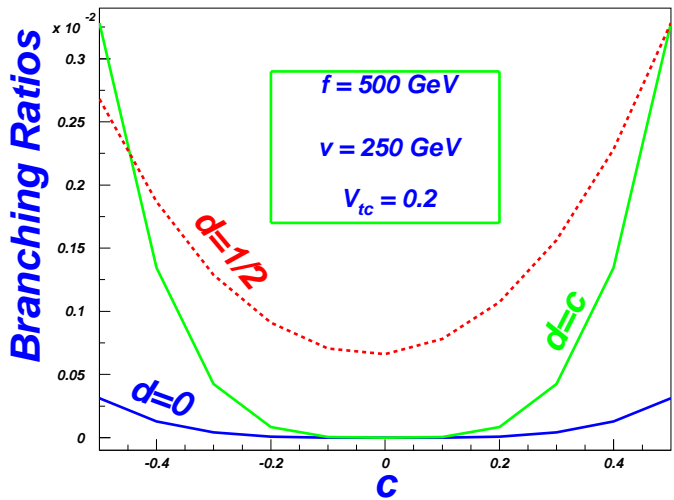}
  \vspace{-0.7cm}\caption[]{ The one-loop level branching ratios of the three processes in the MTH model versus to the structure parameter $c$. }
\label{fig3}
\end{figure}

To see the effect of the structure parameters $c$ and $d$ in Eq.(\ref{tc_coup_gene}) on the branching ratio of $t\to cg$,
we in Fig.\ref{fig3} vary $c$ between $(-\frac{1}{2},\frac{1}{2})$ with $d=0,~c,~\frac{1}{2}$, respectively,
and we can find that the
of course, $c$ and $d$ cannot be equal to zero simultaneously, because if so the coupling $\rho \bar q_i t$ will vanish,
and so does the $t\to cg $ branching ratio.
Normally, they should be the $\pm\frac{1}{2}$ and $0$ (asynchronously).


\def\figsubcap#1{\par\noindent\centering\footnotesize(#1)}
 \begin{figure}[htb]%
 \begin{center}       \hspace{-1.5cm}
  \parbox{5cm}{\epsfig{figure=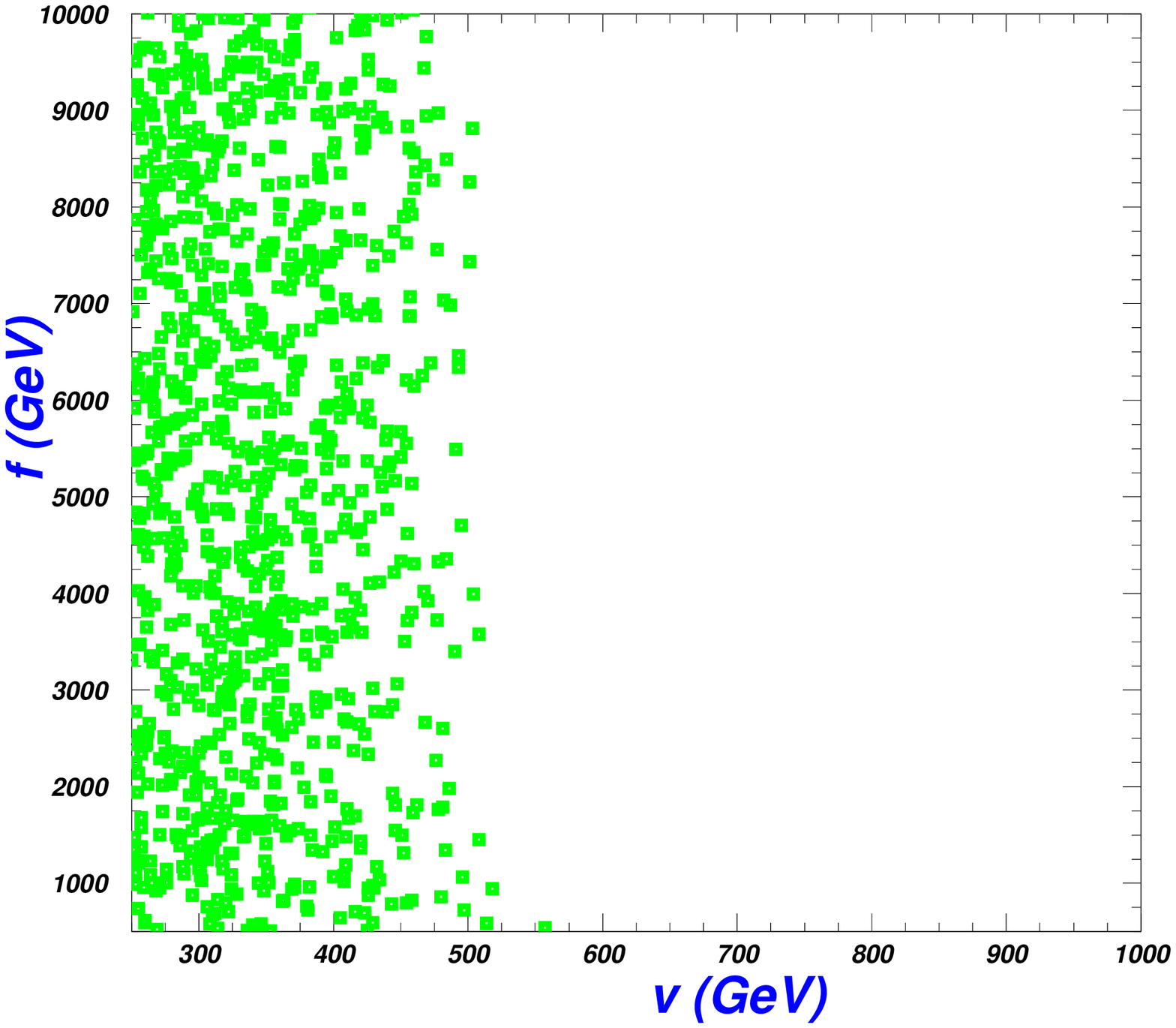,width=5cm}\figsubcap{a}}
  \parbox{5cm}{\epsfig{figure=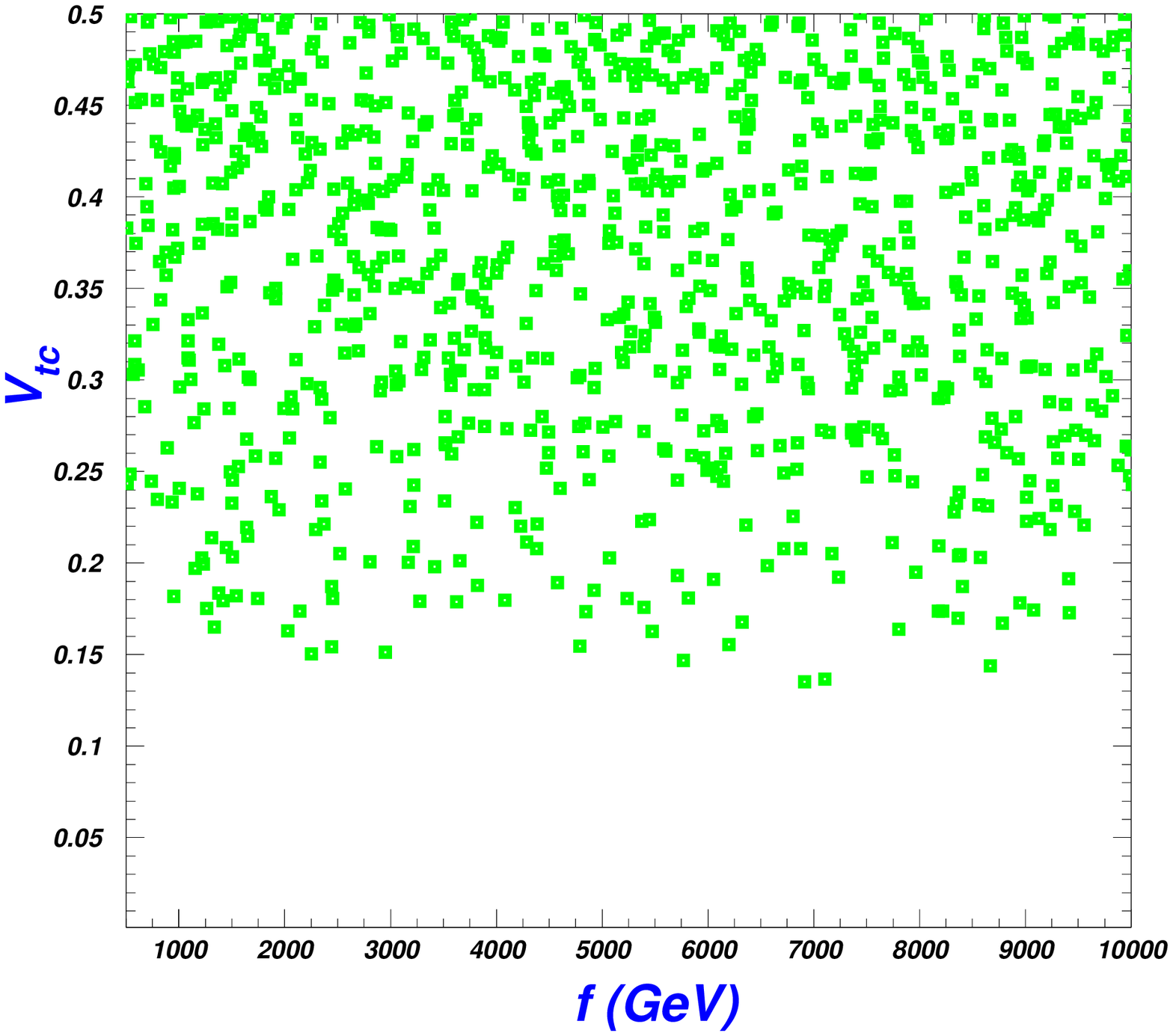,width=5cm}    \figsubcap{b}}
  \parbox{5cm}{\epsfig{figure=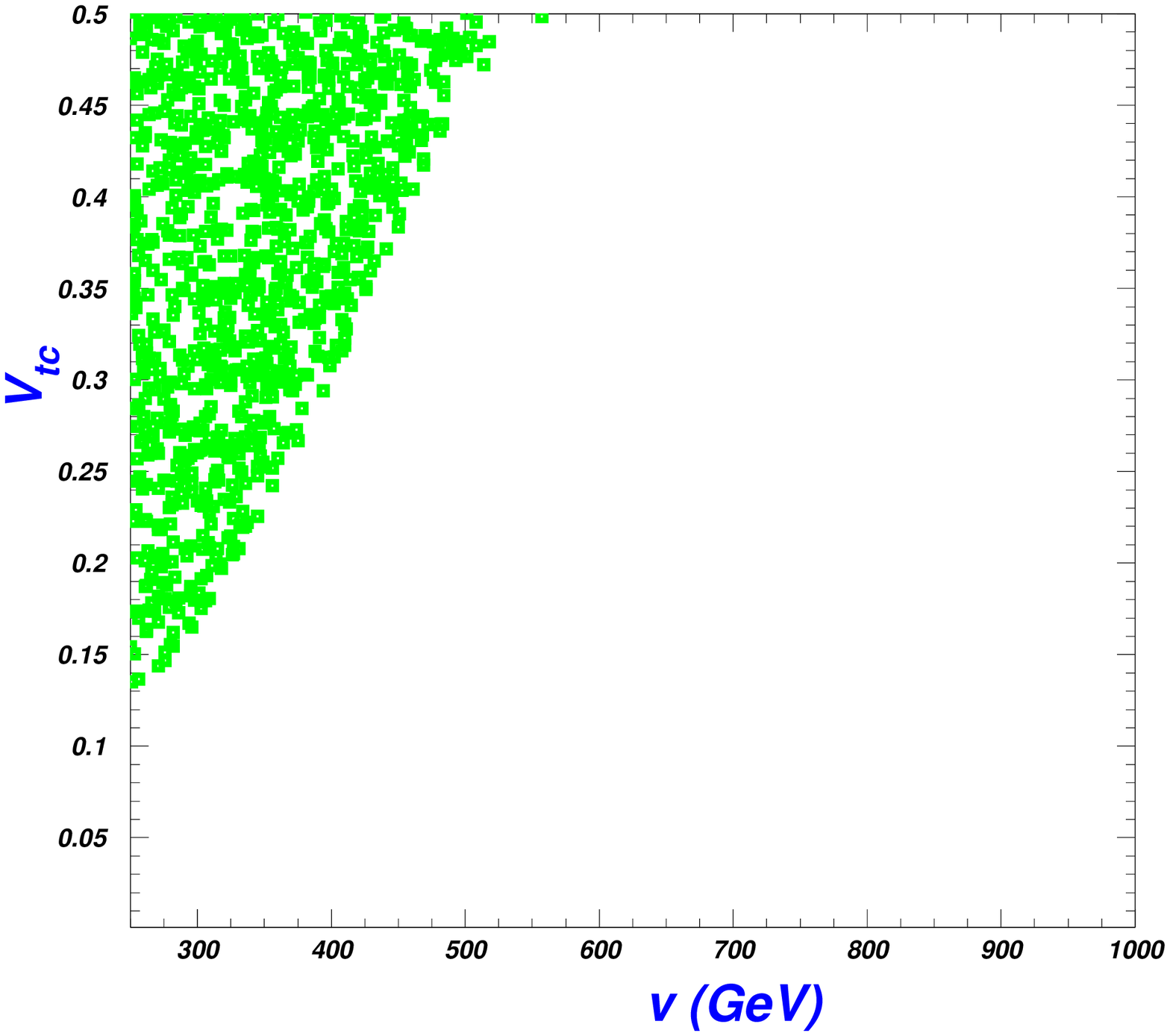,width=5cm}    \figsubcap{c}}
 \caption{ The $t\to cg$ detective contours on $v\sim f$, $f\sim V_{tc}$, and $v\sim V_{tc}$
 in the MTH model.  \label{fig4} }
   \end{center}
 \end{figure}
In Fig.\ref{fig2} and  \ref{fig3}, the parameter values are taken to be in the optimum case, i.g, we take
$v=250$ GeV, $f=500$ GeV, and $V_{tc}=0.2$, quite a large coupling.
But actually, the parameters may be in unfavorable case, which is not
possible to be detected at the colliders, so
we scan the whole parameter spaces to find out the possible room to be accessible to the model.
If it is difficult to mark the signal, the parameters of the models will face severe constraints.
That is, if we cannot find the process,
it may serve as a robust measurement to constrain the model parameters, especially $V_{tc}$.

Fig.\ref{fig4} scans the possibility of the $t\to cg$ branching ratios in light of the detectable level in the following
three parameter ranges:
$500 \leq f \leq 10000$ GeV, $ 250 \leq v \leq 1000$ GeV, and $0.001 \leq V_{tc}\leq 0.5$.
From Fig.\ref{fig4}(a), we can see that the branching ratios prefer to a small $v$, while
insensitive to $f$, which is because that, via Eq.(\ref{vew_v}),
the coupling $\lambda_t=\sqrt{2}m_t/(f sin\vartheta)=\frac{\sqrt{2}m_t}{f sin [v/(\sqrt{2}f)]}$,
so the smaller $v$ is, the larger the branching ratio is.
In the meantime, with two compelling $f$s in the denominator, we can conclude that $v$ contributes
larger than $f$, which is also seen clearly in the next two figures.

However, $V_{tc}$ is close to linear in the amplitude, so the branching ratios will increase rapidly with
the increasing $V_{tc}$. Hence, to arrive at the detectable level, the larger $V_{tc}$ will be preferable,
which can be seen clearly in Fig.\ref{fig4}(b)(c).

From Fig.\ref{fig4}, we can see that the parameters are constrained
in a very narrow space, if the FCNC decay $t\to cg$ cannot be detected.
Since $v$ and $f$ contribute small, $V_{tc}$ is strongly restricted.
 When $V_{tc}\leq 0.13$, the $t\to cg$ branching ratio is normally smaller than the detectable level.

Since it is impossible for the branching ratio of the $t\to c\gamma$ to arrive at the limit Eq.(\ref{tcr_lhc}),
the choosing areas of the parameters $v$, $f$, $V_{tc}$ are not affected by this decay.
As for the $t\to cZ$, there are very little points can arrive at the limit Eq.(\ref{tcz_lhc}),
and the parameter $V_{tc}$ is constrained severely: $V_{tc}\leq 0.45$.

So one can conclude that the MTH model can enhance the branching ratios
of $t\to  c V$  to some significant level from the SM values and may be in the allowed ranges
of the LHC constraints.
\section{Summary and Conclusions}
We have performed a complete one-loop calculation of the flavour-changing top quark decays $(t\to cV)$ ($V=g,\gamma,Z$)
in the context of mirror twin Higgs models.
Since the LHC experimental searches are concerned, performing some searches of FCNC top decays will be possible and
viable.
And some Refs \cite{top-review} have provided the projected limits for higher energies on top FCNCs at the LHC and ILC.
From these data we can see clearly that even in the higher energetic Run-II
of the LHC, the sensitivity will not reach the limit to probe the small branching ratios as obtained
in the theoretical calculations from the SM.
However, there are many BSM scenarios in which these branching ratios are enhanced quite large,
even to the level that may be probed in the Run-II of LHC in some parameter space.
The aim of this work is to look into this issue of rare decays in one of the popular BSM scenario, i.e., MTH models.
We show that all the decay widths of $t\to cV$ do change much from the SM value for favorable parameters
as the consequence of the colorless top partner in this kind of model.
These results are not wield since the coupling between the scalar and the quarks $y_t= m_t/(fsin\vartheta)$
can be quite large in some parameter spaces. And with the development of the future collider,
the decays may be much more hopeful to be probed than nowadays.

{\it Acknowledgements:} Guo-Li Liu would like to thank Fei Wang for very helpful discussions.
This work was supported by the National Natural Science Foundation of China
under grant 11675147, 11775012 and by the Academic Improvement Project of Zhengzhou
University.

\end{document}